\def\papertitle{Sound Analysis and Synthesis Adaptive in Time and Two Frequency Bands}
\def\paperauthorA{Marco Liuni}
\def\paperauthorB{Peter Balazs}
\def\paperauthorC{Axel R\"obel}

\documentclass[twoside,a4paper]{article}
\usepackage{dafx_11}
\usepackage{amsmath,amssymb,amsfonts,amsthm}
\usepackage{euscript}
\usepackage[applemac]{inputenc}
\usepackage[T1]{fontenc}
\usepackage{ifpdf}

\usepackage[english]{babel}
\usepackage{caption}
\usepackage{subfig, color}

\ninept

\usepackage{verbatim} 

\newif\ifpdf
\ifx\pdfoutput\relax
\else
   \ifcase\pdfoutput
      \pdffalse
   \else
      \pdftrue
\fi

\ifpdf 
  \usepackage[pdftex,
    pdftitle={\papertitle},
    colorlinks=false, 
    bookmarksnumbered, 
    pdfstartview=XYZ 
  ]{hyperref}
\usepackage[pdftex]{graphicx}
  \usepackage[figure,table]{hypcap}
\else 
  \usepackage[dvips]{epsfig,graphicx}
  \usepackage[dvips,
    colorlinks=false, 
    bookmarksnumbered, 
    pdfstartview=XYZ 
  ]{hyperref}
  \usepackage[figure,table]{hypcap}
\fi

\usepackage{graphicx}

\usepackage{caption}

\usepackage{latexsym}
\usepackage{mathrsfs}
\usepackage{psfrag}

\usepackage{times}

\title{\papertitle}

\newcommand{\m}{\mathbb}

\newcommand{\om}{\omega}

\newcommand{\de}{\mathrm{d}}

\newcommand{\beq}{\begin{equation}}
\newcommand{\eeq}{\end{equation}}
\newcommand{\lsc}{\langle}
\newcommand{\rsc}{\rangle}

\threeaffiliations{
\paperauthorA, \sthanks{This work was supported by grants from region Ile de France}}
{\href{http://anasynth.ircam.fr/home/}{UMR STMS  IRCAM - CNRS - UPMC} \\ Paris, France \\
{\tt \href{mailto:marco.liuni@ircam.fr}{marco.liuni@ircam.fr}}
}
{\paperauthorB, \sthanks{This work was partially supported by  by  the WWTF project MULAC ('Frame Multipliers: Theory and Application in Acoustics; MA07-025) }}
{\href{http://www.kfs.oeaw.ac.at}{Acoustics Research Institute} \\ Austrian Academy of Sciences \\ Vienna, Austria\\
{\tt \href{mailto:peter.balazs@oeaw.ac.at}{peter.balazs@oeaw.ac.at}}
}
{\paperauthorC,}
{\href{http://anasynth.ircam.fr/home/}{UMR STMS  IRCAM - CNRS - UPMC} \\ Paris, France \\
 {\tt \href{mailto:axel.roebel@ircam.fr}{axel.roebel@ircam.fr}}
}

\begin{document}
\ifpdf 
  \DeclareGraphicsExtensions{.png,.jpg,.pdf}
\else  
  \DeclareGraphicsExtensions{.pdf}
\fi

\maketitle

\begin{abstract}
We present an algorithm for sound analysis and resynthesis
with local automatic adaptation of time-frequency resolution. There exists several algorithms allowing to adapt the analysis window depending on its time or frequency location; in what follows we propose a method which select the optimal resolution depending on both time and frequency. We consider an approach that we denote as \emph{analysis-weighting}, from the point of view of Gabor frame theory. We analyze in particular the case of different adaptive time-varying resolutions within two complementary frequency bands; this is a typical case where perfect signal reconstruction cannot in general be achieved with fast algorithms, causing a certain error to be minimized. We provide examples of adaptive analyses of a music sound, and outline several possibilities that this work opens.
\end{abstract}

\section{Introduction}
\label{sec:intro}

Traditional analysis methods based on single sets of atomic functions offer limited possibilities concerning the variation of the resolution. Moreover, the optimal analysis parameters are often set depending on an a-priori knowledge of the signal characteristics. Analyses with a non-optimal resolution result in a blurring or sometimes even a loss of information about the original signal, which affects every kind of later treatment: visual representation, features extraction and processing among others. This motivates the research for adaptive methods, conducted at present in both the signal processing and the applied mathematics communities: they lead to the possibility of analyses whose resolution locally change according to the signal features.\\

We present an algorithm 
with local automatic adaptation of time-frequency resolution. 
In particular, we use \emph{nonstationary Gabor frames} \cite{JBD09} of windows with compact time supports, being able to adapt the analysis window depending on its time or frequency location. 
For compactly supported windows fast reconstruction algorithms are possible, see \cite{JBD09,nsdgt10,ltfatnote15}: all along the paper we will indicate as \emph{fast} a class of algorithms whose principal computational cost is due to the Fourier transform of the signal.\\

In the present paper we want to go a step beyond and adapt the window in time {\em and} frequency. This case has been detailed in \cite{Do10} among others.
This can be possible, and frame theory \cite{Ch03} would help in providing perfect reconstruction synthesis methods (if no information is lost).  
 However, this is a typical case where the calculation of the dual frame for the signal reconstruction cannot in general be achieved with a fast algorithm: 
 thus a choice must be done between a slow analysis/re-synthesis method guaranteeing perfect reconstruction and a fast one giving an approximation with a certain error. There are, at least, two interesting approaches to obtain fast algorithms:
\begin{itemize}
\item \textbf{filter bank}: 
the signal is first filtered with an invertible bank of $P$ pass band filters, to obtain $P$ different band limited signals; for each of these bands a different nonstationary Gabor frame $\{g_{k,l}^p\}$ of windows with compact time support is used, with $g_k^p$ the time-dependent window function. 
The other members of the frame are time-frequency shifts of  $g_k^p$,
\beq\label{eq:frame_def} g_{k,l}^p = g_k^p(t - a^p_k) e^{2\pi i b^p_k l t} ~,\eeq
where $k,~l \in \m{Z}$ and $a^p_k,b^p_k$ are the time location and frequency step associated to the $p$-th frame at the time index $k$. We will write NGF to indicate a nonstationary Gabor frame in the time case, and we will always assume to be in the painless case \cite{Da86}. Each band-limited signal is perfectly reconstructed with an expansion of the analysis coefficients in the dual frame $\{\widetilde{g_{k,l}}^p\}$.
Note that by this notation we denote the dual frame for a fixed $p$.
 By appropriately combining the reconstructed bands we obtain a perfect reconstruction of the original signal. An important remark is that the reconstruction at every time location is perfect as long as all the frequency coefficients within all the $P$ analyses are used. On the other hand, for every analysis we are interested in considering only the frequency coefficients corresponding to the considered band, thus introducing a reconstruction error. 
\item \textbf{analysis - weighting}: the signal is first analyzed with $P$ NGFs $\{g_{k,l}^p\}$ of windows with compact time support . Each analysis is associated to a certain frequency band, and its coefficients are weighted to match this association. We look for a reconstruction formula to minimize the reconstruction error when expanding the weighted coefficients within the union of the $P$ individual dual frames  $\cup_{p=1}^P \{\widetilde{g_{k,l}}^p\}$.

\end{itemize}
We focus here on the second approach, in the basic case of two bands; so we split the frequency dimension into high and low frequencies, with $P = 2$. We provide the algorithm for an automatic adaptation routine: in each frequency band, the best resolution is defined through the optimization of a sparsity measure deduced from the class of \emph{R\'enyi entropies} \cite{BF01}.
As for the filter bank approach, the results detailed in \cite{ME10} indicate a useful solution: they give an exact upper bound of the reconstruction error when reconstructing a compactly supported and essentially band-limited signal from a certain subset of its analysis coefficients within a Gabor frame.\\ 

In the first section, the analysis-weighting method is treated with an extension of the weighted Gabor frames approach \cite{BAG10}, which will give us a closed reconstruction formula. The second section is dedicated to the sparsity measures we use for the automatic adaptation, with an insight on how weighting techniques of the analysis coefficients can lead to measures with specific features. We then close the paper with some examples and an overview on the perspectives of our research.

\section{Reconstruction from Weighted Frames}
\label{sec:peter} 

Let $P\in \m{N}$ and $\{g_{k,l}^p\}$ be different NGFs, $p= 1,\ldots,P$, where $k$ and $l$ are the time and frequency location, respectively.
We will consider weight functions $0\leq w^p( \nu ) \leq \infty$: for every $p$, they only depend on the frequency location. 
The idea is to smoothly set to zero the coefficients not belonging to the frequency portion which the $p$-th analysis has been assigned to; in this way, every analysis will just contribute to the reconstruction of the signal portion of its pertinence, so high or low frequencies respectively when $P = 2$.
For each NGF $\{g_{k,l}^p\}$ we write $c_{k,l}^p = w^p(b_k^p l) \lsc f,g_{k,l}^p \rsc$ to indicate the weighted analysis coefficients, and we consider the following reconstruction formula: 
\beq\label{form:peter_rec} \widetilde{f} = \mathscr{F}^{-1} \left( \frac{1}{\mathrm{p}(\nu)} \mathscr{F}\left(\sum_{p=1}^P \sum_{k,l} \mathrm{r}(p,k,l) \right) \right)~,\eeq
where $\mathrm{p}(\nu) = \sharp \{p:w^p(\nu) \geq \epsilon \}~$ and for every $\epsilon > 0$, $\mathrm{r}(p,k,l)$ is 0 if $w^p(b_k^p l) < \epsilon$, else
\beq\label{form:weight_rec} \mathrm{r}(p,k,l) =  \left( w^p(b_k^p l) \lsc f,g_{k,l}^p \rsc  \right)\frac{1}{w^p(b_k^p l)} \widetilde{g_{k,l}}^p~.\eeq

We see that non-zero weights cancel each other: this reconstruction formula still makes sense, as the goal is exactly to find a reconstruction as an expansion of the $c_{k,l}^p$.\\

We give now an interpretation of the introduced formula.
If $w^p$ is a semi-normalized sequence for each $p$, that is there exist constants $m_p$ and $n_p$ such that $0 < m_p \leq w^p(b_k^p l) \leq n_p$
and $\epsilon \le m_p$ $\forall p$, then $\mathrm{p}(\nu) = p$ and the equation \eqref{form:peter_rec} becomes
\beq\label{form:peter_rec_2} \widetilde{f} =  \frac{1}{P} \sum_{p=1}^P \sum_{k,l}  \left( w^p(b_k^p l) \lsc f,g_{k,l}^p \rsc  \right)\frac{1}{w^p(b_k^p l)} \widetilde{g_{k,l}}^p = f~.\eeq
This is related to the concept of weighted frames detailed in \cite{BAG10}, as in the hypothesis of semi-normalization the sequence $w^p(b_k^p l) g_{k,l}^p$ is a frame with $\frac{1}{w^p(b_k^p l)} \widetilde{g_{k,l}}^p$ as one of its dual.
For weights which are not bounded from below, but still non-zero, the reconstruction still works: the sequences $w^p(b_k^p l) \cdot g_{k,l}^p$ are not frames anymore (for each $p$), but complete Bessel sequences (also known as upper semi-frames \cite{AP09}). This reconstruction can be unstable, though.

In our case, these hypotheses are not verified, as we need to set to zero a certain subset of the coefficients within both of the analyses; thus the equation \eqref{form:peter_rec} will in general give an approximation of $f$. In section \ref{sec:ex_w} we give an example of reconstruction following this approach, evaluating the reconstruction error; further theoretical and numerical examinations should be realized, as we are interested to find an upper bound for the error depending on:
\begin{itemize}
\item the signal spectral features at frequencies $\nu$ where  $\mathrm{p}(\nu) > 1~;$
\item the features of the $w^p$ sequences and the $\mathrm{p}(\nu)$ function.
\end{itemize}

A first natural choice for the weights $w^p$ is a binary mask; first because this is the worst case in terms of reconstruction error, as we are multiplying in the frequency domain with a rectangular window before performing an inverse Fourier transform. Thus the analysis of the error with a binary masking establish a bound to the error obtained with a smoother mask. Moreover, with a binary mask the reconstruction formula takes the very simple form detailed in equation \eqref{form:peter_rec_3}, allowing a direct implementation derived from the general full band algorithm. So we consider $P=2$ and $\om_c$ a certain cut value, then 

\begin{displaymath} w^1( \nu ) = \left\{ \begin{array}{ll}
1 & \textrm{if $\nu \leq \om_c$}\\ 0 & \textrm{if $\nu > \om_c$}
\end{array} \right. \vspace{-7mm} \end{displaymath}
\beq\label{spec_models}\eeq
\vspace{1mm}

\noindent and $w^2(\nu) = 1 - w^1(\nu) $. In this case $\mathrm{p}(\nu) = 1$ for every frequency $\nu$ and the equation \eqref{form:peter_rec} becomes
\beq\label{form:peter_rec_3}  \widetilde{f} =  \sum_{b_k^p l \leq \om_c} \lsc f,g_{k,l}^1 \rsc \widetilde{g_{k,l}}^1 + \sum_{b_k^p l > \om_c} \lsc f,g_{k,l}^2 \rsc \widetilde{g_{k,l}}^2~.\eeq
 The reconstruction error in this case will in general be large at frequencies corresponding to coefficients close to the cut value $\om_c$; we envisage that a way to reduce this error is to allow the $w^p$ weights to have a smooth overlap; this results in more coefficients form different analyses contributing to the reconstruction of a same portion of signal, thus weakening their interpretation.

\section{R\'enyi Entropy Evaluation of Weighted Spectrograms}
\label{sec:renyi}

The representation we take into account is the spectrogram of a signal $f$: it is the squared modulus of the Short-Time Fourier Transform (STFT) of $f$ with  window $g$,
which is defined by 
\beq\label{eq:def_stft} {\mathcal V}_{g} f \left( u, \xi \right) = \int f (t) \overline{g(t - u)} e^{- 2 \pi i \xi t} \de t~,\eeq 
and so the spectrogram is  $\mathrm{PS}_f (t , \om) = \left| {\mathcal V}_g f (t , \om) \right|^2$.  Given a Gabor frame $\{g_{k,l}\}$ we obtain a sampling of the spectrogram coefficients considering $z_{k,l} = |\lsc f,g_{k,l} \rsc|^2$. With an appropriate normalization, both the continuous and sampled spectrogram can be interpreted as probability densities. The idea to use Rnyi entropies as sparsity measures for time-frequency distributions has been introduced in \cite{BF01}: minimizing the complexity or information of a set of time-frequency representations of a same signal is equivalent to maximizing the concentration, peakiness, and therefore the sparsity of the analysis. Thus we will consider as \emph{best} analysis the sparsest one, according to the minimal entropy evaluation.\\

Given a signal $f$ and its spectrogram $\mathrm{PS}_f$, the \emph{Rnyi entropy} of \emph{order} $\alpha > 0,~\alpha \neq 1$ of $\mathrm{PS}_f$ is defined as follows
\beq\label{ren_ent_def}  \mathrm{H}_{\alpha}^R(\mathrm{PS}_f) =\frac{1}{1-\alpha}~\log_2 \iint_R \bigg(\frac{\mathrm{PS}_f (t,\om)}{\iint_R\mathrm{PS}_f(t',\om')\de t'\de\om'}\bigg)^{\alpha}\de t \de \om~, \eeq
where $R\subseteq \m{R}^2$ and we omit its indication if equality holds. Given a discrete spectrogram obtained through the Gabor frame $\{g_{k,l}\}$, we consider $R$ as a rectangle of the time-frequency plane $R = [t_1,t_2]\times[\nu_1,\nu_2] \subseteq \m{R}^2$. It identifies a sequence of points $G$ on the sampling grid defined by the frame. As a discretization of the original continuous spectrogram, every sample $|z_{k,l}|^2$ is related to a time-frequency region of area $ab$, where  $a$ and $b$ are respectively the time and frequency steps; we thus obtain the discrete Rnyi entropy measure directly from \eqref{ren_ent_def},
\beq\label{ren_ent_disc} \mathrm{H}_{\alpha}^G[\mathrm{PS}_f ] =  \frac{1}{1-\alpha}\log_2 \sum_{k,l\in G} \bigg(\frac{z_{k,l}}{\sum_{[k',l']\in G} z_{k',l'}}\bigg)^{\alpha}  + \log_2(ab)~.\eeq

We consider now another weight function $0\leq w(k,l) \leq \infty$; instead of weighting the STFT coefficients $\lsc f,g_{k,l} \rsc$ as we did in Section \ref{sec:peter}, we weight here the discrete spectrogram obtaining a new distribution $z^*_{k,l} = w(k,l) z_{k,l}$ which is not necessarily the spectrogram of a signal: nevertheless, by the definition of $w(k,l)$, its Rnyi entropy can still be evaluated from \eqref{ren_ent_disc}. This value gives an information of the concentration of the distribution within the time-frequency area emphasized by the specific weight function: as we show in section \ref{sec:ex_adap}, this can be useful for the customization of the adaptation procedure.\\

We will focus on discretized spectrograms with a finite number of coefficients, as dealing with digital signal processing requires to work with finite sampled signals and distributions. As $\alpha$ tends to one this measure converges to the Shannon entropy, which is therefore included in this larger class. General properties of Rnyi entropies can be found in \cite{Re61}, \cite{BS93} and \cite{Zy04}; in particular, given $P$ a probability density, $\mathrm{H}_{\alpha}(P)$ is a non increasing function of $\alpha$, so $\alpha_1 < \alpha_2 \Rightarrow \mathrm{H}_{\alpha_1}(P)\geq \mathrm{H}_{\alpha_2}(P)~.$ Moreover, for every order $\alpha$ the Rnyi entropy $\mathrm{H}_{\alpha}$ is maximum when $P$ is uniformly distributed, while it is minimum and equal to zero when $P$ has a single non-zero value. As we are working with finite discrete densities we can also consider the case $\alpha = 0$ which is simply the logarithm of the number of elements in $p$; as a consequence $\mathrm{H}_0[p] \geq \mathrm{H}_{\alpha}[p]$ for every admissible order $\alpha$. As long as we can give an interpretation to the $\alpha$ parameter, this class of measures offers a largely more detailed information about the time-frequency representation of the signal.\\

\subsection{Adaptive procedure}

We choose a finite set $S$ of admissible scaling factors, and realize different scaled version of a window $g$, 
\beq\label{scaling} g^{s}(t) = \frac{1}{\sqrt{s}}~g\bigg(\frac{t}{s}\bigg)~, \eeq
so that the discretized temporal support of the scaled windows $g^s$ still remains inside $G$ for any $s\in S$. In our case, $G$ is a rectangle with the time segment analyzed as horizontal dimension and the whole frequency lattice as vertical: at each step of our algorithm, this rectangle is shifted forward in time with a certain overlap with the previous position. By fixing an $\alpha$, the sparsest local analysis is defined to be the one with minimum Rnyi entropy: thus the optimization is performed on the scaling factor $s$, and the best window is defined consequently, with a similar approach to the one developed in \cite{jaill1}. With the weight functions introduced above, we are also able to limit the frequency range of the rectangle $G$ at each time location: adaptation is thus obtained over the time dimension for each weighted spectrogram, so in our case for each frequency band enhanced. An interpolation is performed over the overlapping zones to avoid abrupt discontinuities in the tradeoff of the resolutions: in the examples given in section \ref{sec:algo_ex}, the spectrogram segment for the entropy evaluation includes four spectrogram frames of the largest window, and the overlapping zone corresponds to three frames of the largest window. The temporal sizes of the segment and the overlap are deduced accordingly.\\
The time-frequency adapted analysis of the global signal is finally realized by opportunely assembling the slices of local sparsest analyses obtained with the selected windows.

\subsection{Biasing spectral coefficients through the $\mathbb{\alpha}$ parameter}
\label{ssec:sub_alpha}
\begin{figure}[h!t]

\begin{minipage}[b]{1.0\linewidth}
  \centering
  \centerline{\includegraphics[width=9.5cm]{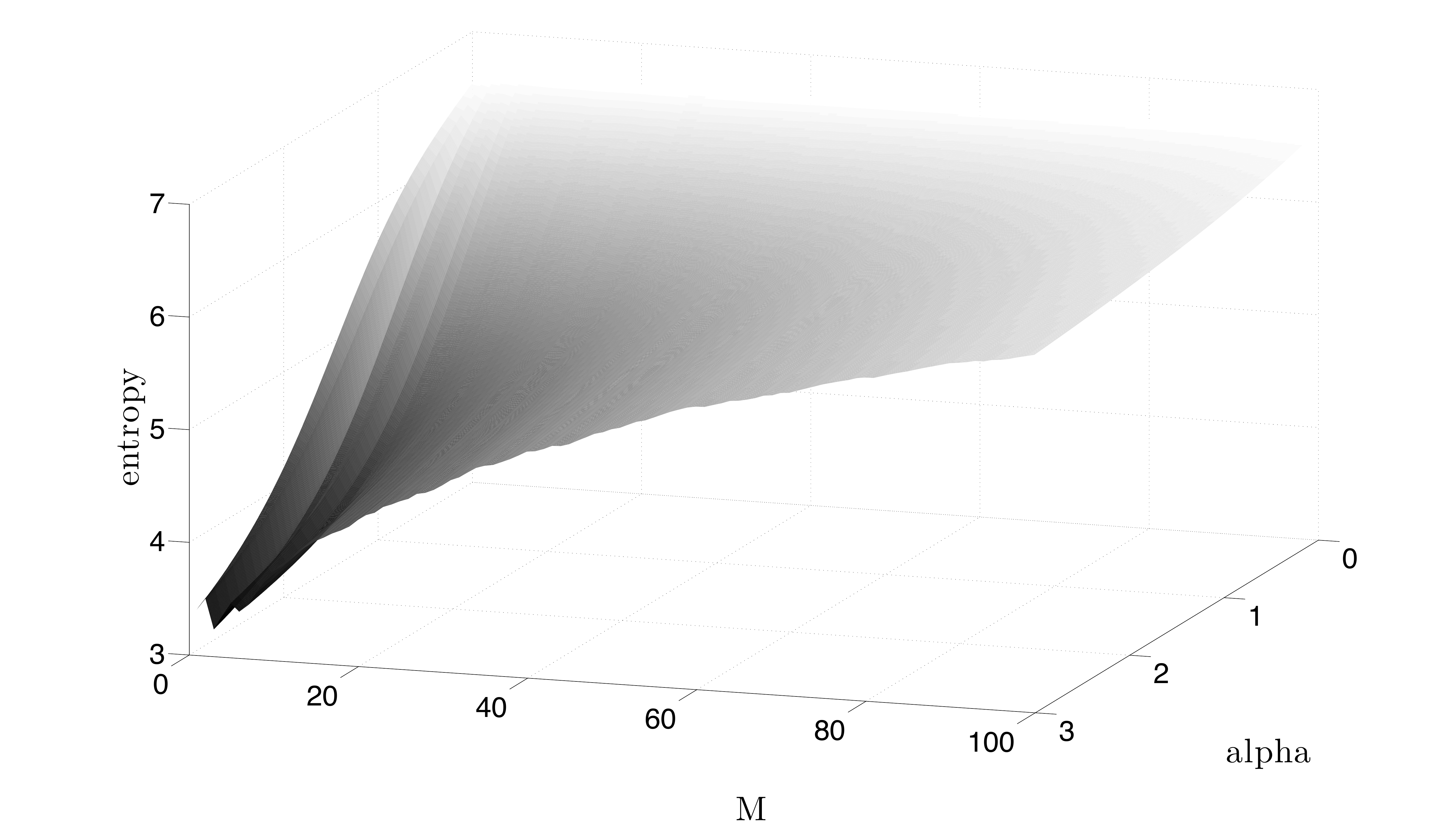}}\medskip
\vspace{-0.5cm}
\end{minipage}

\caption{Rnyi entropy evaluations of the $D_M$ vectors with varying $\alpha$; the distribution becomes flatter as $M$ increases. Therefore increasing $\alpha$ favors a sparse representation (see text). }
\label{fig:test_alpha}
\end{figure}

The $\alpha$ parameter in equation \eqref{ren_ent_def} introduces a biasing on the spectral coefficients; to have a qualitative description of this biasing, we first consider a collection of simple spectrograms composed by a variable amount of large and small coefficients. We realize a vector $D$ of length $N = 100$ generating numbers between 0 and 1 with a normal random distribution; then we consider the vectors $D_M,~1\leq M\leq N$ such that

\begin{displaymath} D_M[k] = \left\{ \begin{array}{ll}
D[k] & \textrm{if $k\leq M$}\\ \frac{D[k]}{20} & \textrm{if $k > M$}
\end{array} \right. \vspace{-8mm} \end{displaymath}
\beq\label{spec_models}\vspace{5mm}\eeq
and then normalize to obtain a unitary sum. We then apply Rnyi entropy measures with $\alpha$ varying between 0 and 3: these are the values that we use to adopt for music signals. As we see from figure \ref{fig:test_alpha}, there is a relation between the number of large coefficients $M$ and the slope of the entropy curves for the different values of $\alpha$.
For $\alpha = 0$, $\mathrm{H}_0[D_M]$ is the logarithm of the number of non-zero coefficients and it is therefore constant; when $\alpha$ increases, we see that densities with a small amount of large coefficients gradually decrease their entropy, faster than the almost flat vectors corresponding to larger values of $M$. This means that by increasing $\alpha$ we emphasize the difference between the entropy values of a peaky distribution and that of a nearly flat one. The sparsity measure, we consider, selects as best analysis the one with minimal entropy, so reducing $\alpha$ rises the probability of less peaky distributions to be chosen as sparsest: in principle, this is desirable as weaker components of the signal, such as partials, have to be taken into account in the sparsity evaluation.\\ 

\begin{figure}[h!t]

\begin{minipage}[b]{1.0\linewidth}
  \centering
  \centerline{\includegraphics[width=9.5cm]{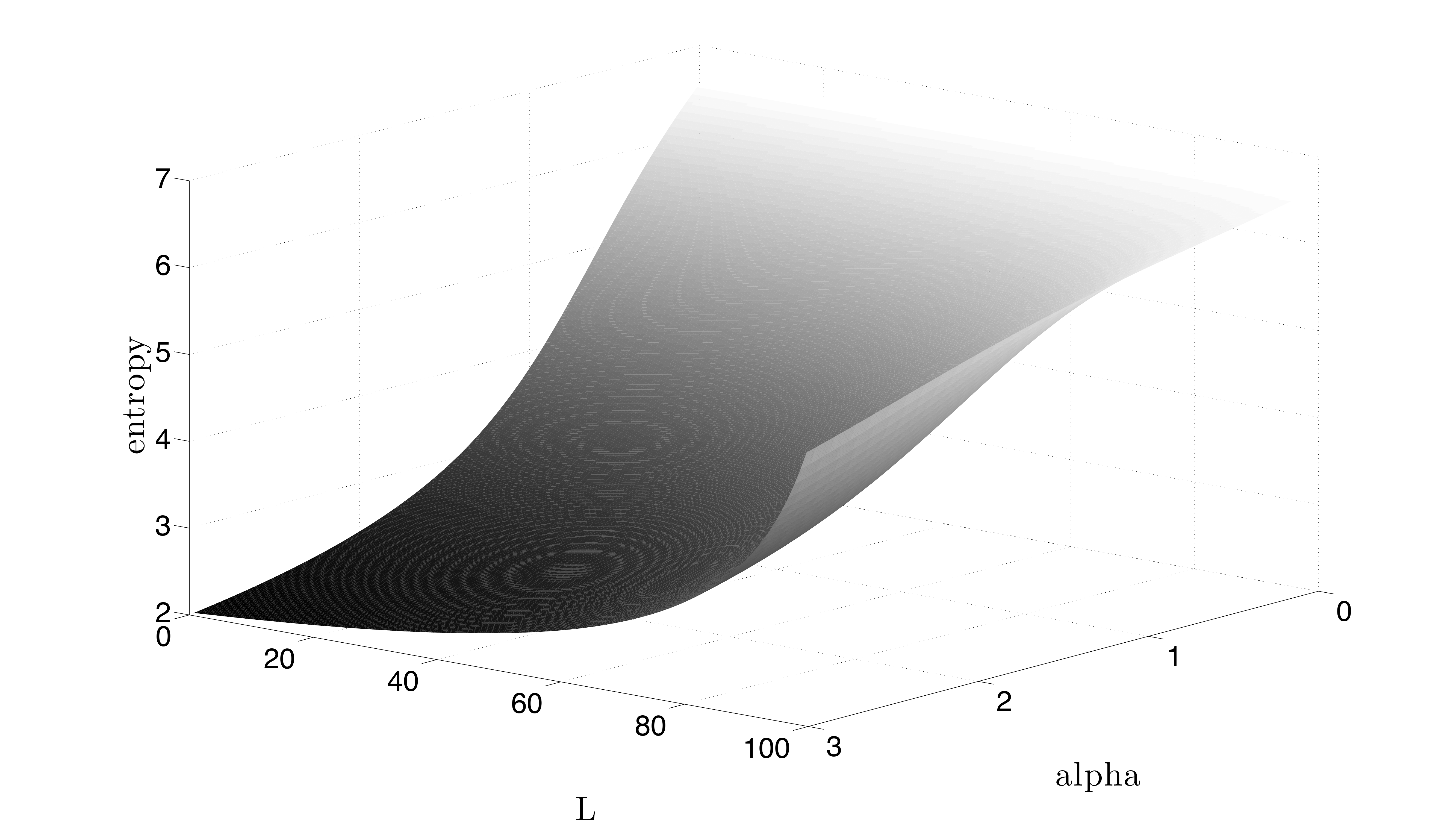}}\medskip
\vspace{-0.5cm}
\end{minipage}

\caption{Rnyi entropy evaluations of the $D_L$ vectors with varying $\alpha$, $N_{part} = 5 $ and $R_{part} = 2$; the entropy values rise differently as $L$ increases, depending on $\alpha$: this shows that the impact of the noise level on the entropy evaluation depends on the entropy order (see text).}
\label{fig:test_alpha_2}
\end{figure}

The second example we consider shows that the just mentioned principle should be applied with care, as a small coefficient in a spectrogram could be determined by a partial as well as by a noise component; with an extremely small $\alpha$, the best window selected could vary without a reliable relation with spectral concentration, depending on the noise level within the sound. We illustrate how noise has to be taken in account when tuning the $\alpha$ parameter by means of another model of spectrogram: taking the same vector $D$ considered previously, and two integers $1 \leq N_{part}$, $1 \leq R_{part}$, we define $D_L$ like follows:

\begin{displaymath} D_L[k] = \left\{ \begin{array}{lll}
1 & \textrm{if $k = 1$} \vspace{2mm} \\ 
\frac{D[k]}{R_{part}} & \textrm{if $1 < k \leq N_{part}$}\vspace{2mm} \\
\frac{D[k]}{R_{noise}} & \textrm{if $ k > N_{part}~$.}
\end{array} \right. \vspace{-12mm} \end{displaymath}
\beq\label{spec_models_2}\vspace{8mm}\eeq
where $R_{noise} = \frac{R_{part}}{L},~L \in [\frac{1}{16},1]$; then we normalize to obtain a unitary sum. This vectors are a simplified model of the spectrograms of a signal whose coefficients correspond to one main peak, $N_{part}$ partials with amplitude reduced by $R_{part}$ and some noise whose amplitude varies, proportionally to the $L$ parameter, from a negligible level to the one of the partials. Applying Rnyi entropy measures with $\alpha$ varying between 0 and 3, we obtain the figure \ref{fig:test_alpha_2}, which shows the impact of the noise level $L$ on the evaluations with different values of $\alpha$.

The increment of $L$ corresponds to a strengthening of the noise coefficients, causing the rise of the entropy values for any $\alpha$. The key point is the observation of how they rise, depending on the $\alpha$ value: the convexity of the surface in figure \ref{fig:test_alpha_2} increases as $\alpha$ becomes larger, and it describes the impact of the noise level on the evaluation; the stronger convexity when $\alpha$ is around 3 denotes an higher robustness, as the noise level needs to be high to determine a significant entropy variation. Our tests show that, as a drawback, in this way we lower the sensitivity of the evaluation to the partials, and the measure keeps almost the same profile for every $R_{part} > 1$.\\
On the other hand, when $\alpha$ tends to 0 the entropy growth is almost linear in $L$, showing the significant impact of noise on the evaluation, as well as a finer response to the variation of the partials amplitude. As a consequence, the tuning of the $\alpha$ parameter has to be performed according to the desired tradeoff between the sensitivity of the measure to the weak signal components to be observed, and the robustness to noise. In our experimental experience, the value of 0.7 is appropriate for both speech and music signals.

\section{Algorithms and examples}
\label{sec:algo_ex}

We give here two examples of the methods described above: the first shows an application of two different weights on the spectrogram of a given sound, which determines two different choices for the optimal resolutions; the second is a reconstruction with the algorithm detailed in Section \ref{sec:peter}.


\subsection{Adaptation with Different Masks}
\label{sec:ex_adap}

We can privilege a certain subset of the analysis coefficients to drive the adaptation routine, instead of considering them all with the same importance. 
For example, the adaptation within the $p$-th band could be determined from the coefficients laying at a certain small distance from the band central frequency.\\

\begin{figure}
\begin{minipage}[b]{1.0\linewidth}
  \centering
  \centerline{\includegraphics[width=10.5cm]{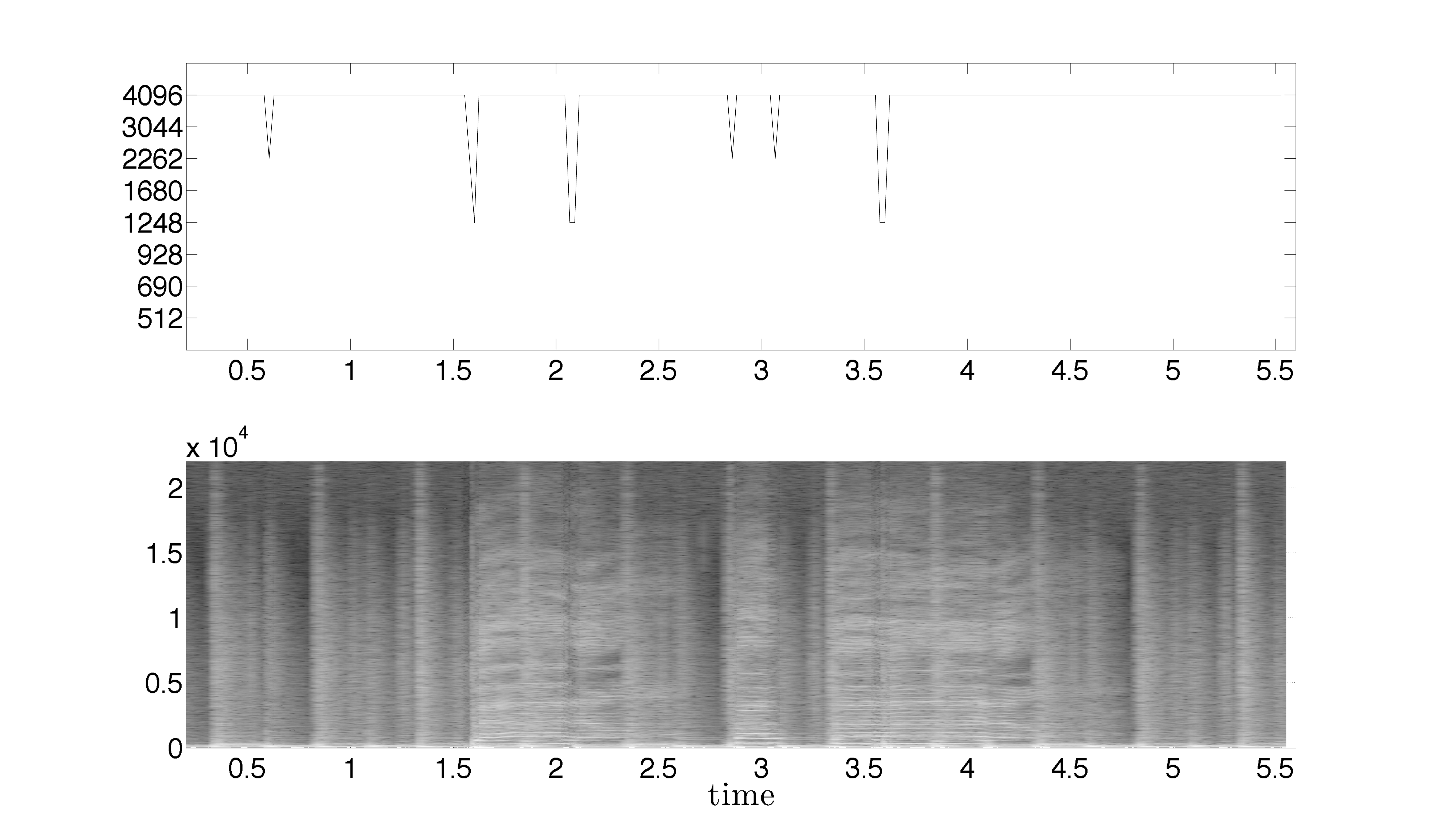}}\medskip
\vspace{-0.5cm}
\end{minipage}

\caption{Adaptive analysis with a mask privileging frequencies below 300Hz, on a music signal with a bass guitar, a drum set and a female singing voice starting from second 1.54: on top, best window size chosen as a function of time; at the bottom, adapted spectrogram of the analyzed sound file.}
\label{fig:mask_low}
\end{figure}

\begin{figure}

\begin{minipage}[b]{1.0\linewidth}
  \centering
  \centerline{\includegraphics[width=10.5cm]{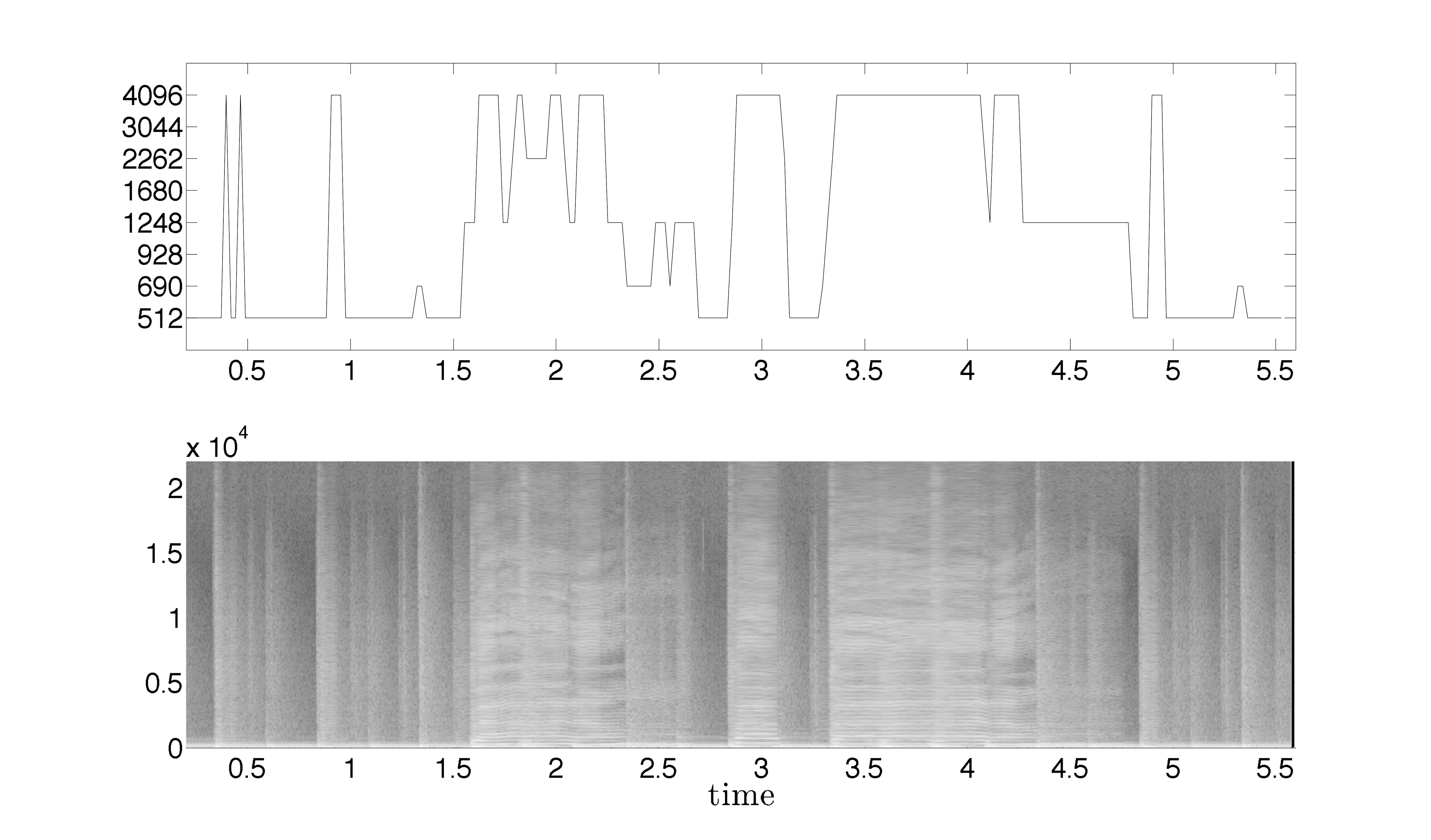}}\medskip
\vspace{-0.5cm}
\end{minipage}

\caption{Adaptive analysis with a mask privileging frequencies above 300Hz, on a music signal with a bass guitar, a drum set and a female singing voice starting from second 1.54: on top, best window size chosen as a function of time; at the bottom, adapted spectrogram of the analyzed sound file.}
\label{fig:mask_high}
\end{figure}

Figures \ref{fig:mask_low} and \ref{fig:mask_high} are realized with an improved version of the algorithm described in \cite{LRRR10}, which allows for a weighting of the analysis coefficients which concerns only the adaptation routine, and not the analysis and re-synthesis. Thus, we obtain different adapted analyses depending on the frequency area we wish to privilege, still preserving perfect reconstruction: the sound we analyze is a music signal with a bass guitar, a drum set and a female singing voice starting from second 1.54. We use two different complementary binary masks, the first setting to zero the spectrogram coefficients corresponding to frequencies higher than 300Hz, the second doing the opposite. As we can see in Figure \ref{fig:mask_low}, with the first mask we obtain an analysis where the largest window is privileged; this is the best frequency resolution for the bass guitar sound, which is prominent in the considered band. The only points where shorter windows are chosen correspond to strong transients, as bass or voice attacks, where the time precision is enhanced.\\
With the second mask, low frequencies are ignored in the adaptation step, and as a consequence we obtain a different optimal analysis: the smallest window is generally selected, yielding an higher time resolution which is best adapted to the percussive sounds; moreover, we see that the largest window is chosen corresponding to the presence of the singing voice, whose higher harmonics belong to the considered band and determine a better frequency resolution to be privileged.\\ 

In both cases we calculate the difference between the signal reconstructed and the original one; we use a 16 bit audio file, whose amplitude is represented in the range $[-1,1]$ with double precision: the maximum absolute value of the differences between corresponding time samples, as well as the root mean square error over the entire signal, are both of order $10^{-16}$.

\subsection{Analysis-Weighting Example}
\label{sec:ex_w}

\begin{figure}
\begin{minipage}[b]{1.0\linewidth}
  \centering
  \centerline{\includegraphics[width=10cm]{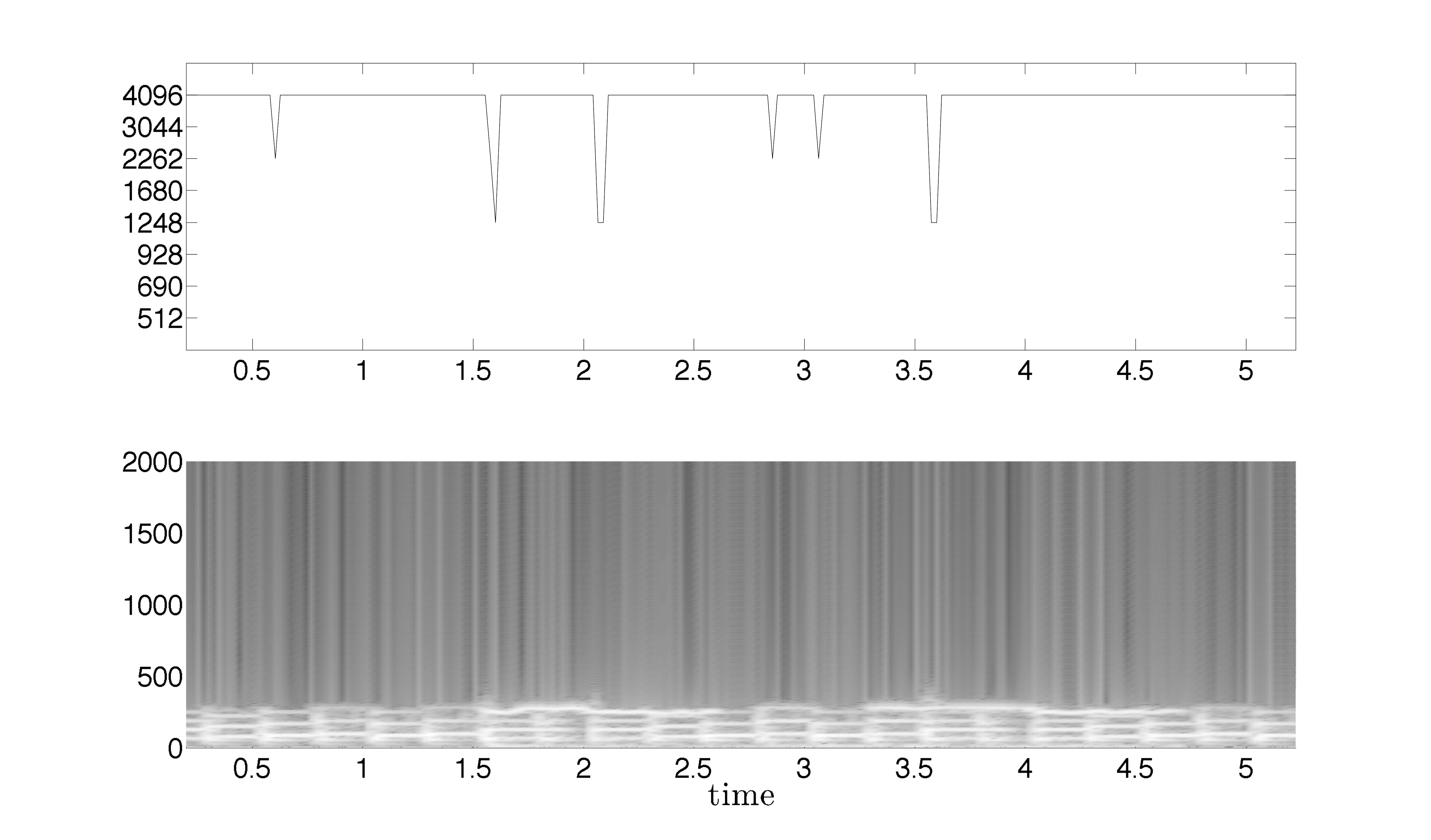}}\medskip
\vspace{-0.5cm}
\end{minipage}

\caption{Low-frequencies reconstruction from the masked adapted analysis of a music signal with a bass guitar, a drum set and a female singing voice starting from second 1.54: on top, best window size chosen as a function of time. At the bottom, spectrogram of the analyzed band with a 4096 samples Hamming window, 3072 samples overlap and 4096 frequency points; the frequency axis is bounded to 2kHz to focus on the reconstructed region.}
\label{fig:ex_low}
\end{figure}

\begin{figure}
\begin{minipage}[b]{1.0\linewidth}
  \centering
  \centerline{\includegraphics[width=10cm]{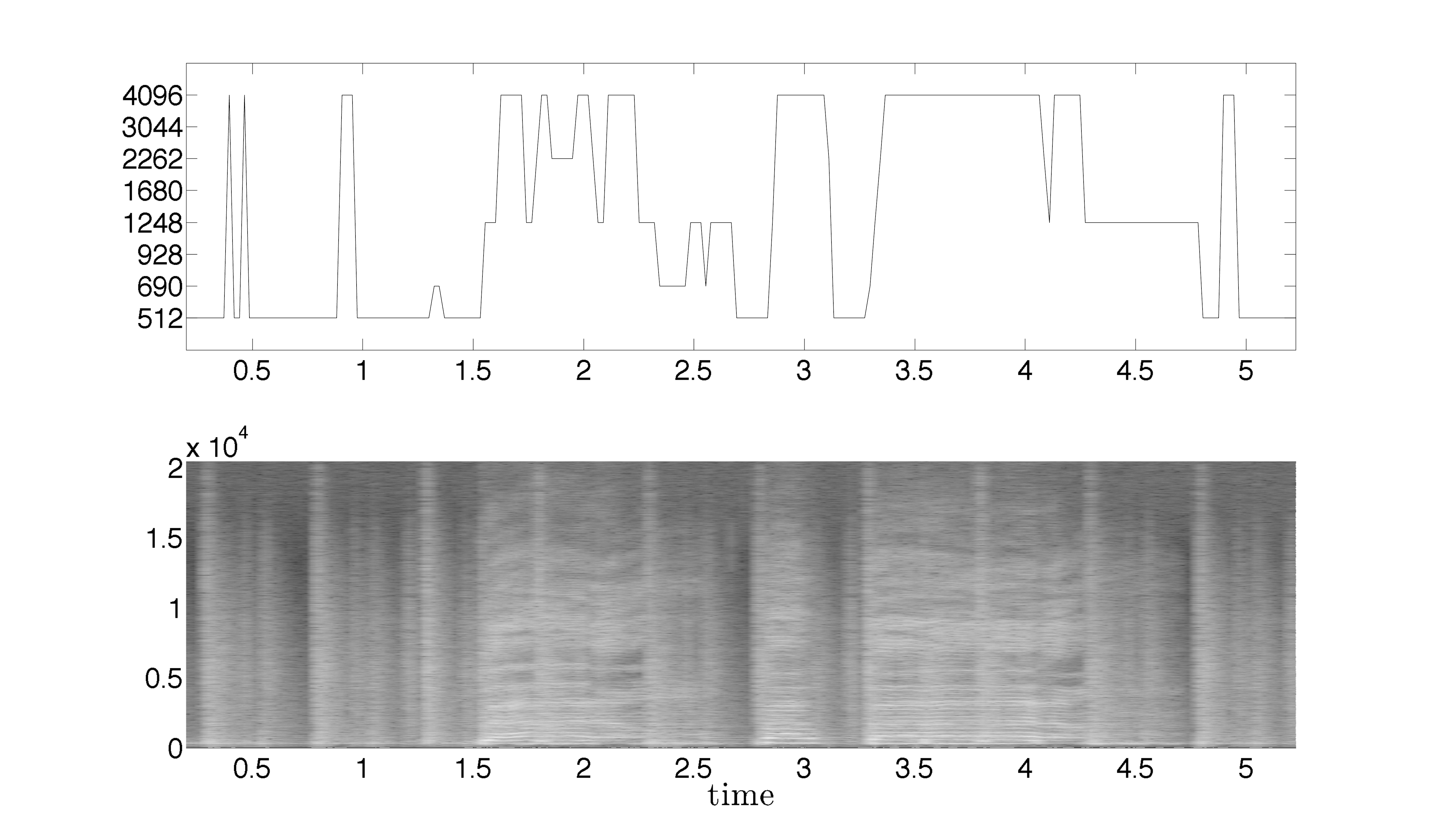}}\medskip
\vspace{-0.5cm}
\end{minipage}

\caption{High-frequencies reconstruction from the masked adapted analysis of a music signal with a bass guitar, a drum set and a female singing voice starting from second 1.54: on top, best window size chosen as a function of time; at the bottom, spectrogram of the analyzed band with a 4096 samples Hamming window, 3072 samples overlap and 4096 frequency points.}
\label{fig:ex_high}
\end{figure}

We show here an example of the approximation of a signal applying the formula \eqref{form:peter_rec_3}, within the analysis-weighting approach using a binary mask: as detailed in Sections \ref{sec:peter} and \ref{sec:renyi}, we analyze a signal with different stationary Gabor frames; the sound we consider is the same of the section \ref{sec:ex_adap}, and the binary mask is still obtained with a cut frequency of 300Hz, while the sampling rate is 44.1kHz. We modify the coefficients of all these analyses with the mask $w^1(\nu)$, and build the NGF $\{g_{k,l}^1\}$ with resolutions adapted to the low frequencies optimizing the entropy of the masked analyses. Then we repeat this step with the mask $w^2(\nu)$ and build the NGF $\{g_{k,l}^2\}$. We finally calculate the duals of the two NGFs, which can be done in these cases with fast algorithms, and re-synthesize the two signal bands: for these examples, the reconstruction is performed with the SuperVP phase vocoder by Axel R\"obel \cite{svp}.\\
Figure \ref{fig:ex_low} shows the spectrogram of the lower signal band, reconstructed with the low-frequencies adapted analysis. This spectrogram is computed with a fixed window, which is the largest one within the set considered; the choice of the best window is given as well, to give information about how the reconstruction is performed at each time. Figure \ref{fig:ex_high} is obtained in the same way, considering the upper band reconstruction. The approximation of the original sound is then given by the sum of the two bands.\\

\begin{figure}
\begin{minipage}[b]{1.0\linewidth}
  \centering
  \centerline{\includegraphics[width=10cm]{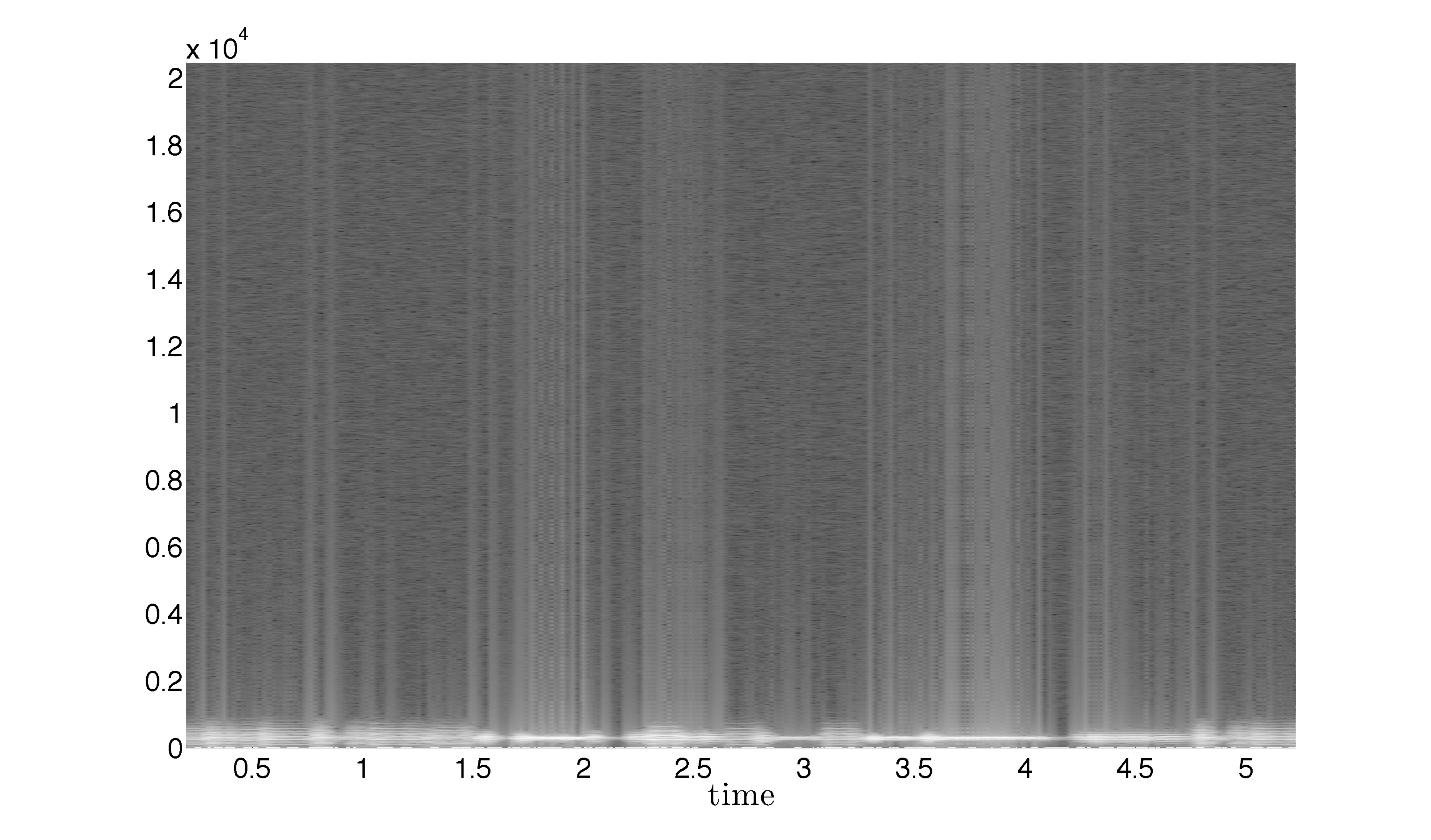}}\medskip
\vspace{-0.5cm}
\end{minipage}

\caption{Spectrogram of the reconstruction error given by the described method on a music signal with a bass guitar, a drum set and a female singing voice starting from second 1.54; spectrogram obtained with a 4096 samples Hamming window, 3072 samples overlap and 4096 frequency points.}
\label{fig:ex_er}
\end{figure}

The reconstruction error we obtain is higher than the one in the previous examples: the maximum absolute value of the samples differences is  0.0568, while the root mean square error is 0.0099. With the choice of a binary mask, the only way to reduce the error is to set the cut frequency in a range where the signal energy is low: unfortunately, music signals generally do not have large low-energy bands; moreover, the interest of our method relies in the possibility for the cut frequency to be variable, in order to freely select the adaptation criterium.\\ 
Figure \ref{fig:ex_er} shows the spectrogram of the difference between the original sound and the reconstructed one, and we see that the spectral content of the error is concentrated at the cut frequency. The alteration introduced has negligible perceptual effects, so that the original signal and the reconstruction are hard to be distinguished: this aspect needs to be quantified; when dealing with the approximation of music signals, the objective error measures do not give any information about the perceptual meaning of the error. The accuracy of a method has thus to be evaluated by means of measures taking into account the human auditory system as well as listening tests.\\ 

Another element to consider is the overlap between the weight functions introduced in section \ref{sec:peter}: if we allow them for an overlap over a sufficiently large frequency band, we envisage that the error would be reduced. The sense of this point can be clarified considering the causes of the reconstruction error: windows with compact time support cannot have a compactly supported Fourier transform; from the analysis point of view, this means that a spectrogram coefficient affects the signal reconstruction among the whole frequency dimension. We can limit such an influence with a choice of well-localized time-frequency atoms: even if their frequency support is not compact, they have a fast decay outside a certain region. If we cut with a binary mask outside a certain band, the reconstruction error comes mainly from the fact that we are setting to zero the contribution of atoms whose Fourier transforms spread into the band of interest: if the atoms are well-localized, only a few of them actually have an impact.\\
Formula \eqref{form:peter_rec} gives an ideal reference: if the overlap is the entire frequency dimension, weights are non-zero, thus we have a perfect reconstruction from the weighted coefficients. When some weights are zero and weight functions do overlap, the normalization factor in the formula \eqref{form:peter_rec} is greater than one in the overlapping frequency interval. This reduces the impact of the errors coming from individual re-syntheses: on the other hand, the fact of summing them all imposes a limit to the achievable global error reduction.\\ 
A further improvement of this formula is to put different weights at the denominator in \eqref{form:peter_rec_2}, with an effective amplification or reduction of the contributions coming from individual coefficients. To keep the perfect reconstruction valid in the case of semi-normalized norms, a possibility is to obtain the different weights as a function of the analysis weights depending also on the overlap.

\section{Conclusions and perspectives}
\label{sec:persp}

We have sketched the first steps of a promising research project about the local automatic adaptation of time-frequency sound representations: a first question which arises is how to display a representation of the signal such the one described; there are two possibilities involving weighted means of the coefficients at a certain time-frequency location:
\begin{itemize}
\item  $d_{k,l} = \frac{1}{\sum_p w^p} \cdot \sum \limits_p c_{k,l}^p$, displaying $\left| d_{k,l}\right|$, or 
\item $d^{(A)}_{k,l} = \frac{1}{\sum_p w^p} \cdot \sqrt{ \sum \limits_p \left| c_{k,l}^p \right|^2}$.
\end{itemize} 
In a previously proposed method \cite{LRRR10} the algorithm keeps the original coefficients in memory; with this approach, we can use the reconstruction scheme mentioned in \eqref{form:peter_rec_4}. A further new question would be how to reconstruct the signal from an expansion of the $d_{k,l}$ or $d^{(A)}_{k,l}$ coefficients. Straightforward numerical examples could give some numerical insights.

If $d^{(A)}_{k,l}$ is used, we also have to address the problem of the phase.
This approach is useful when dealing with spectrogram transformations where the phase information is lost, as with reassigned spectrogram or spectral cepstrum. We could either use an iterative approach, like the one described in \cite{GL84} adapted to frame theory, or use a system with a high redundancy (see \cite{BRCPED06}).\\

From a computational point of view, we are interested in limiting the size of the signal for the direct and inverse Fourier transforms in \eqref{form:peter_rec}, as this will largely improve the efficiency of the algorithm. A different form of the formula \eqref{form:peter_rec} in this sense is
\beq\label{form:peter_rec_4} \widetilde{f} = \sum_{p,k,l} c_{k,l}^p  \mathscr{F}^{-1}\left( \frac{1}{\mathrm{p}(\nu)}  \mathscr{F}\left(\frac{\widetilde{g_{k,l}}^p}{w^p(b_k^p l)}\right)\right) \eeq
whose properties have to be further investigated.

Later we would also investigate the properties of time-variant filters by multiplying these new sets of coefficients, resulting in new kinds of frame multipliers \cite{xxlmult1}.
Using an optimized way to analyze acoustical signal, will, therefore, also lead to a better control of such adaptive filters.

\bibliographystyle{IEEEbib}
\bibliography{dafx11_bib} 

\begin{thebibliography}{10}

\bibitem{JBD09}
F.~Jaillet, P.~Balazs, M.~D{\"o}rfler, and N.~Engelputzeder,
\newblock ``Nonstationary {G}abor {F}rames,''
\newblock in {\em Proc. of SAMPTA'09}, Marseille, France, May 18-22, 2009.

\bibitem{nsdgt10}
P.~Balazs, M.~D\"orfler, N.~Holighaus, F.~Jaillet, and G.~Velasco,
\newblock ``Theory, {I}mplementation and {A}pplications of {N}onstationary
  {G}abor {F}rames,''
\newblock submitted,
  \href{http://www.univie.ac.at/nonstatgab/pdf\_files/badohojave11\_04042011.p%
df}{http://www.univie.ac.at/nonstatgab/pdf\_files/}
  \mbox{badohojave11\_04042011.pdf}, 2011.

\bibitem{ltfatnote15}
P.~L. S\o{ndergaard}, B.~Torr\'esani, and P.~Balazs,
\newblock ``The {L}inear {T}ime {F}requency {A}nalysis {T}oolbox,''
\newblock
  \href{http://www.univie.ac.at/nuhag-php/ltfat/toolboxref.pdf}{http://www.uni%
vie.ac.at/nuhag-php/ltfat/toolboxref.pdf}.

\bibitem{Do10}
{M}onika {D}{\"o}rfler,
\newblock ``{Q}uilted frames - a new concept for adaptive representation,''
\newblock {\em {A}dvances in {A}pplied {M}athematics, to appear}, 2010,
\newblock
  \href{http://arxiv.org/pdf/0912.2363}{http://arxiv.org/pdf/0912.2363}.

\bibitem{Ch03}
O.~Christensen, Ed.,
\newblock {\em An Introduction To Frames And Riesz Bases},
\newblock Birkh\"auser, Boston, Massachussets, USA, 2003.

\bibitem{Da86}
I.~Daubechies A. Grossmann~Y. Meyer,
\newblock ``Painless nonorthogonal expansions,''
\newblock {\em J. Math. Phys.}, vol. 27, pp. 1271--1283, May 1986.

\bibitem{BF01}
R.G. Baraniuk P. Flandrin A.J.E.M. Janssen~O.J.J. Michel,
\newblock ``Measuring {T}ime-{F}requency {I}nformation {C}ontent {U}sing the
  {R}\'enyi {E}ntropies,''
\newblock {\em IEEE Trans. Info. Theory}, vol. 47, no. 4, pp. 1391--1409, May
  2001.

\bibitem{ME10}
E.~Matusiak Y.~C. Eldar,
\newblock ``Sub-{N}yquist sampling of short pulses: Part i,''
\newblock http://arxiv.org/abs/1010.3132v1.

\bibitem{BAG10}
P.~Balazs, J.-P. Antoine, and A.~Gribo{\'s},
\newblock ``Weighted and controlled frames: mutual relastionships and first
  numerical properties,''
\newblock {\em Int. J. Wav, Mult. Info. Proc.}, vol. 8, no. 1, pp. 109--132,
  2010.

\bibitem{AP09}
J.-P. Antoine and P.~Balazs,
\newblock ``Frames and semi-frames.,''
\newblock to appear in Journal of Physics A: Mathematical and Theoretical.
  \href{http://arxiv.org/pdf/1101.2859v2}{http://arxiv.org/pdf/1101.2859v2}.

\bibitem{Re61}
A.~R\'enyi,
\newblock ``On {M}easures of {E}ntropy and {I}nformation,''
\newblock in {\em Proc. Fourth Berkeley Symp. on Math. Statist. and Prob.},
  Berkeley, California, June 20-30, 1961, pp. 547--561.

\bibitem{BS93}
F.~Schl\"ogl C.~Beck, Ed.,
\newblock {\em Thermodynamics of chaotic systems},
\newblock Cambridge University Press, Cambridge, Massachusetts, USA, 1993.

\bibitem{Zy04}
K.~Zyczkowski,
\newblock ``R\'enyi {E}xtrapolation of {S}hannon {E}ntropy,''
\newblock {\em Open Systems \& Information Dynamics}, vol. 10, no. 3, pp.
  297--310, Sept. 2003.

\bibitem{jaill1}
F.~Jaillet and B.~Torr{\'e}sani,
\newblock ``Time-frequency jigsaw puzzle: adaptive and multilayered {G}abor
  expansions,''
\newblock {\em International Journal for Wavelets and Multiresolution
  Information Processing}, vol. 1, no. 5, pp. 1--23, 2007.

\bibitem{LRRR10}
M.~Liuni A. R\"obel M. Romito~X. Rodet,
\newblock ``A reduced multiple {G}abor frame for local time adaptation of the
  spectrogram,''
\newblock in {\em Proc. of DAFx10}, Graz, Austria, September 6-10, 2010, pp.
  338 -- 343.

\bibitem{svp}
Axel R\"obel,
\newblock ``Super{VP},''
\newblock
  \href{http://anasynth.ircam.fr/home/software/supervp}{http://anasynth.ircam.%
fr/home/} \mbox{software/supervp}.

\bibitem{GL84}
D.W. Griffin~J.S. Lim,
\newblock ``Signal {E}stimation from {M}odified {S}hort-{T}ime {F}ourier
  {T}ransform,''
\newblock {\em IEEE Trans. Acoust. Speech Signal Process.}, vol. 32, no. 2, pp.
  236--242, Apr. 1984.

\bibitem{BRCPED06}
{R}adu {B}alan, {P}ete {C}asazza, and {D}an {E}didin,
\newblock ``{O}n signal reconstruction without phase.,''
\newblock {\em {A}ppl. {C}omput. {H}armon. {A}nal.}, vol. 20, no. 3, pp.
  345--356, 2006.

\bibitem{xxlmult1}
P.~Balazs,
\newblock ``Basic definition and properties of {B}essel multipliers,''
\newblock {\em Journal of Mathematical Analysis and Applications}, vol. 325,
  no. 1, pp. 571--585, January 2007.

\end{thebibliography}

\end{document}